\documentclass{article}
\pdfoutput=1

\usepackage{arxiv}

\usepackage[utf8]{inputenc} 
\usepackage[T1]{fontenc}    
\usepackage{hyperref}       
\usepackage{url}            
\usepackage{booktabs}       
\usepackage{amsfonts}       
\usepackage{nicefrac}       
\usepackage{microtype}      
\usepackage{bm}             
\usepackage{amsmath}
\usepackage{graphicx}
\usepackage{colortbl}

\usepackage[
backend=biber,
style=nature,
sorting=none
]{biblatex}
\graphicspath{ {./images/} }

\title{Individual-level Modeling of COVID-19 Epidemic Risk}

\newcommand{\Lik}{\mathcal{L}}
\newcommand{\Sus}{\mathcal{S}}
\newcommand{\Exp}{\mathcal{E}}
\newcommand{\Inf}{\mathcal{I}}
\newcommand{\Rec}{\mathcal{R}}

\author{
 Andres Colubri \\
  Broad Institute of MIT and Harvard\\
  Cambridge, MA 02142, USA \\
  \texttt{andres@broadinstitute.org} \\
  \And
 Kailash Yadav \\
  Elucidata\\
  New Delhi, Delhi 110049, India \\
  \texttt{kailash.yadav@elucidata.io} \\
  \And  
 Abhishek Jha \\
  Elucidata\\
  Cambridge, MA 02139, USA \\
  \texttt{abhishek.jha@elucidata.io} \\  
   \And
   Pardis Sabeti \\
   Broad Institute of MIT and Harvard\\
   Cambridge, MA 02142, USA \\
   \texttt{pardis@broadinstitute.org} \\
}

\begin{document}
\maketitle

\begin{abstract}
The ongoing COVID-19 pandemic calls for a multi-faceted public health response comprising complementary interventions to control the spread of the disease while vaccines and therapies are developed. Many of these interventions need to be informed by epidemic risk predictions given available data, including symptoms, contact patterns, and environmental factors. Here we propose a novel probabilistic formalism based on Individual-Level Models (ILMs) that offers rigorous formulas for the probability of infection of individuals, which can be parameterised via Maximum Likelihood Estimation (MLE) applied on compartmental models defined at the population level. We describe an approach where individual data collected in real-time is integrated with overall case counts to update the a predictor of the susceptibility of infection as a function of individual risk factors.
\end{abstract}


\section{Introduction}
The COVID-19 Pandemic \cite{Velavan2020, Fauci2020} has emerged as the most serious health crisis that humanity has faced since the 1918 Influenza Pandemic \cite{Viboud2018}. Its causal pathogen, SARS-CoV-2 \cite{Lu2020}, is a coronavirus new to the human population with unique molecular \cite{Wrapp2020}, physiopathological \cite{Zou2020}, and epidemiological \cite{Liu2020} features. This has resulted in the exponential spread of COVID-19 around the world, with over 22 million confirmed cases and nearly 800,000 deaths worldwide as reported by the COVID-19 interactive web dashboard from John Hopkins University (https://coronavirus.jhu.edu/map.html), at the time of this writing \cite{Dong2020}.

As part of the public interventions that aim to reduce the transmission of COVID-19, precautionary self-isolation of the general population and quarantining of suspected and confirmed mild cases is a strategy that can substantially reduce the effective reproductive number of the disease, $R_e$ \cite{Wang2020}. This has the important consequence of "flattening the epidemic curve" until herd immunity is achieved, either by infection or vaccination, and therefore avoid overwhelming the health care system \cite{Kissler2020}. Active monitoring of contacts via traditional contact tracing by health care workers \cite{eames2003}, potentially complemented and expanded with proximity-sensing tracking mobile apps \cite{Ferretti2020}, could further help mitigate transmission \cite{Peak2020}.

In all these interventions, the spatio-temporal modeling of the disease is critical to understand risk factors associated with transmission and, through this, adjust the magnitude and timing of the interventions in order to maximize their change of success. In particular, predicting the epidemic risk of individuals to contract the disease over space and time can help to identify subpopulations under increased risk and to inform interventions such as quarantine. Most importantly, being able to promptly identify who, in a system, is at risk of infection during an outbreak is key to the efficient control of the epidemic. However, developing such models is challenging in a situation like the current pandemic, due to the uncertainty in the epidemiological parameters of a novel pathogen and also to the urgency with which interventions and tools are needed.

In this paper, we adopt a individual-level model (ILM) framework that enable us to express the probability of a susceptible individual being infected as a function of their interactions with the surrounding infectious population while also allowing to incorporate the effect of individually-varying risk factors (e.g., age, pre-existing conditions). ILMs are intuitive and flexible due to be expressed in terms of individual interactions \cite{Gibson1997, Keeling2001, Neal2004} but also computationally costly to parameterise, especially in the case of epidemics in large populations. More recent work has shown how to simplify the likelihood calculations to make ILMs more readily applicable in real-life scenarios \cite{Deardon2010}, and also how geographical covariates can be incorporated \cite{Mahsin2020} in order to represent spatial distribution of infection risk over time and model public health interventions that take into account geographical variation.
 
We first apply the formalism for ILMs to derive an expression for the marginal probability of individual risk of infection as a function of parameters with straightforward epidemiological interpretation and initial estimation. We incorporate symptom and other individual-level data to update the risk of infection based on this new information. We then construct a population-level compartmental SIR epidemic model, where the rate of infection can be estimated from the individual-level probabilities given a random sample of individuals front he population. This allows to express the population-level parameters as a function of the individual-level parameters, and to use partially observed data (overall case counts, and individual risk factors and contacts) to apply Maximum Likelihood Estimate (MLE) within a Partially Observed Markov Process (POMP) framework. The POMP framework enables us to solve a computationally more tractable MLE problem thanks to iterated filtering, an efficient computational method that's based on a sequence of filtering operations which are shown to converge to a maximum likelihood parameter estimate. As result of this approach, we arrive to estimates of individual-level parameters that can be used to predict risk of infection.

The structure of this article is as follows. In Section 2, the formula for individual epidemic risk is derived and applied to evaluate conditional probability on infection given additional individual-level data and the overall likelihood function. In Section 3, the iterated filtering MLE is presented to fit the model parameters. Section 4, applies the iterated filtering MLE algorithm on simulated data generated with an agent-based model (ABM). We conclude the article with a discussion in Section 5.

\section{Model Formalism}

\subsection{Individual-level Models}

We adopt the general ILM formalism described by Deardon et al \cite{Deardon2010}. In these models, each individual at each time $t$ can be in one of three states: susceptible, infected, or recovered, denoted $S(t)$, $I(t)$, or $R(t)$ respectively. Furthermore, time is discretized so that $t$ represents a time interval $[t, t + 1)$. The probability of a susceptible individual $i$ being infected at time t is expressed as follows:

\begin{equation}
P(i, t)=1-\exp\big[ \big( -\Omega_{S}(i) \sum\limits_{j \in C(i, t)} \Omega_{T}(j) \kappa(i, j, t)\big) - \epsilon(i, t)\big]
\end{equation}

In this expression, the set of infectious individuals who interacted with susceptible $i$ in the time interval $[t, t + 1)$ is denoted as $C(i, t)$. Following the notation in \cite{Deardon2010}, the risk factors of a susceptible individual $i$ becoming infected, and an infected individual $j$ transmitting the disease are represented by the functions $\Omega_{S}(i)$ and $\Omega_{T}(j)$, respectively. The time-dependent infection kernel, $\kappa(i, j, t)$, incorporates pairwise risk factors, such as the occurrence and duration of contacts between $i$ and $j$. The "sparks" term, $\epsilon(i, t)$, can be used to represent other sources of infection, such as zoonotic introductions at a given rate.

This formula is the result of assuming a Poisson infectious process in each time interval $[t, t + 1)$. We count the number $k$ of transmission events between susceptible $i$ and infected $j$, which follow a Poisson distribution $f\left( k \right) = \frac{{e^{ - \lambda } \lambda ^k }}{{k!}}$ with $\lambda$ the rate of transmission. Non-infection from $j$ corresponds to $k=0$, so then $f(k=0)=e^{ - \lambda }$. The rate of transmission from $j$ to $i$ is modeled as the product form $\lambda_{i,j}=\Omega_{S}(i)\Omega_{T}(j) \kappa(i, j)$. Non-infection of $i$ from all infected $j$ follows from independence between these Poisson processes, therefore (ignoring the sparks term):

\begin{center}
$1-P(i, t)= \prod\limits_{j}e^{ - \lambda_{i,j} }$
\end{center}

Formally, the ILMs can be extended to incorporate the effect of spatial-dependent risk factors upon the transmission of infectious disease. The resulting GD-ILMs have the general form \cite{Mahsin2020}:

\begin{equation}
P(i, k, t)=1-\exp\big[ \big( -\Omega_{S}(i, k) \sum\limits_{j \in C(i, t)} \Omega_{T}(j, k) \kappa(i, j, t)\big) - \epsilon(i, k, t)\big]
\end{equation}

where each function is now a function of an area index $k$ ranging over the geographical subdivisions under consideration. However, in the context of this manuscript we will only consider simple ILM without explicit geographical dependencies.

\subsection{Model Covariates}

The functions introduced in the previous section, $\Omega_{S}(i)$ and $\Omega_{T}(j)$ are general and can incorporate covariates that determine susceptibility and transmissibility risks through various functional dependencies. We can start by considering a linear form:

\begin{equation}
\Omega_{S}(i)=a_{0} + a_{1} X_{1}(i) + ... + a_{N} X_{N}(i)
\end{equation}

where $a_0>0$ is a constant susceptibility parameter, and $X_{n}(i)$ are covariates that represent various susceptibility factors to be included in the model (e.g.: age, pre-existing conditions, etc.) Thus, $a_{n}$ is the parameter for the $n$-th individual-level covariate.

The transmisibility function has a similar general form:

\begin{equation}
\Omega_{T}(j)=b_{0} + b_{1} Y_{1}(j) + ... + b_{M} Y_{M}(i)
\end{equation}

where $b_0>0$ is a constant transmisibility parameter, $Y_{m}(i)$ and $b_{m}$ the transmisibility covariates and their corresponding coefficients.

The infection kernel is a function of the varying distance between i and j over the course of the time interval $[t, t + 1)$

\begin{equation}
\kappa(i, j, t)=F(\{d_{i,j}(\tau), \tau \in [t, t+1)\})
\end{equation}

The next sections will provide specific details on these factors and concrete assumptions motivated by what is known about COVID-19.

\subsection{Model Simplifications}

The general forms of the $\Omega_{S}(i)$ and $\Omega_{T}(i)$ functions described above allow us to incorporate an arbitrary number of covariates into the model. Here we propose a very simple initial model. The susceptibility function will depend only on "immunity status" of the susceptible individual. We define the variable $X_1(i)$ to be 1 if individual $i$ is over 65 or is immunosuppressed due to some pre-existing condition, 0 otherwise.  Therefore:

\begin{equation}
\Omega_{S}(i)=a_0 + a_1 X_1(i)
\end{equation}

For the transmissibility function, an important factor determining the potential for an infected individual to pass along the virus seems to be the presence or absence of symptoms. So in this case the binary covariate $Y_1(j)$ takes the values 0 or 1 whether the infected individual is aymptomatic/pre-symptomatic or symptomatic, respectively. So we arrive to the following simplified form:

\begin{equation}
\Omega_{T}(j)=b_0 + b_1 Y_1(j)
\end{equation}

As for the the infection kernel, for the time being is just 1 when $j$ is in $C(i, t)$, the contact set of $i$, which includes all the infectious individuals whim whom $i$ was closer than 2 meters for at least 15 minutes in $[t, t+1)$, and 0 otherwise:

\begin{equation}
\kappa(i, j)=\begin{cases} 
                1 & j \in C(i, t) \\
                0 & $otherwise$ 
             \end{cases}
\end{equation}

Finally, we will adopt a zero sparks term. This assumption is reasonable is transmission mainly due to interactions between individuals, and not through the environment (e.g.: contaminated surfaces or fomites). There is some anecdotal evidence that this might be the case in COVID-19 \cite{Ferretti2020}, but for the time being we just assume $\epsilon(i, t)=0$ to keep the models simple.

With these modeling decisions we reach the following expression for the individual probability of infection:

\begin{equation}
P(i, t)=1-\exp\Big( -\big[a_0 + a_1 X_1(i)] \sum\limits_{j \in C(i, t)} \big[b_0 + b_1 Y_1(j))] \Big)\label{eq:10}
\end{equation}

\subsection{Conditioning on Additional Data}

Formula \eqref{eq:10} gives an expression of the marginal probability $P(i, t)$ of infection of individual $i$ in time interval $[t,t+1)$. This probability depends on a number of individual-level and area-level covariates. However, additional data from the individual $i$ can be used to arrive to an updated risk of infection. In particular, we are interested in the probability of infection over the course of the past $d$ days given this new data, $P(I_{[t-d, t]}|D)$, with $d$ defining an appropriate retrospective window of possible infection. Given that the incubation period of COVID-19 is two weeks, then $d=14$ should be a suitable choice to inform quarantining/testing measures. Applying Bayes Theorem to this probability, we can formally write:

\begin{center}
$P(I_{[t-d, t]}|D)=\frac{P(D|I_{[t-d, t]})P(I_{[t-d, t]})}{P(D)}$
\end{center}

The probability of an infection over the course of the past $d$ days can be expressed as a function of the per-day probabilities of infection:

\begin{center}
$P(I_{[t-d, t]}) = P(I_{t-d}) + P(I_{t-d+1}) + ... + P(I_{t})$
\end{center}

since each event (infection $d$ days ago, $d-1$ days ago, and so on) is independent from each other. Furthermore, infection $n$ day ago implies that infection did not happen until exactly that day, and so:

\begin{center}
$P(I_{t-n})=\prod\limits_{l=0}^{d-n-1}\big[1-P(t-d+l)\big]P(t-n)$
\end{center}

where $P(t')$ is precisely \eqref{eq:10}, with indices for individual $i$ omitted for clarity.

More concretely, if D comprises symptom data self-reported by individual $i$, or $S_i$, we can then write the risk of infection over the past $d$ days for individual $i$ at time $t$ given symptom data $S_i$:

\begin{equation}
R(i, t)=r_s\sum\limits_{n=0}^{d}\prod\limits_{l=0}^{d-n-1}\big[1-P(i, t-d+l)\big]P(i, t-n)
\label{eq:13}
\end{equation}

with the coefficient $r_s$ defined as:

\begin{equation}
r_s=\frac{P(S_i|I_{[t-d, t]})}{P(S_i)}
\end{equation}

This coefficient can be thought of the ratio of the observed symptoms given the fact that the individual was recently infected to the prevalence of those symptoms among the general population, which in general should be greater than 1. For instance, if the observed symptoms are cough and fever, then $r_s$ is a measure of how much prevalent those symptoms are among people infected with COVID-19, and could be estimated from currently available data. In fact, a recent study \cite{Menni2020} looked at the predictive power of symptoms self-reported in the US and UK with the COVID Symptom Study mobile app \cite{COVIDSymptomStudy}. This study presents a logistic regression predictor of infection given a number of symptom predictors:

\begin{equation}
\begin{split}
\ln[p/(1-p)]=1.32 - 0.01 \times \mbox{age} + 0.44 \times \mbox{sex} + 1.75 \times \mbox{smell and taste loss} + \\
0.31 \times \mbox{cough} + 0.49 \times \mbox{fatigue} + 0.39 \times \mbox{skipped meals}
\end{split}
\end{equation}

Where $p=P(I|S)$. We make the assumption that, for a sufficiently long period of time $d$, the conditional probability $P(S|I_{[t-d, t]})$ is simply $P(S|I)$, where $I$ represents infection at some moment in the past. With this assumption in place, we can connect the COVID Symptom Study prediction model with out probabilistic formalist by means of:

\begin{equation}
\frac{P(S|I)}{P(S)}=\frac{P(I|S)}{P(I)}=r_s
\end{equation}

where $P(I)$ would represent the overall prevalence of COVID-19 infection. This allows us to write the following final formula for the risk score:

\begin{equation}
R(i, t)=\frac{P(I|S_i)}{P(I)}\sum\limits_{n=0}^{d}\prod\limits_{l=0}^{n-1}\big[1-P(i, t-d+l)\big]P(i, t-d+n)
\label{eq:14}
\end{equation}

Given knowledge of the individual's symptoms, demographic and medical covariates, and their recent contact history, it would be then possible to calculate this risk of infection.

\section{Maximum likelihood in POMPs}

Following \cite{Deardon2010}, given the $S(t)$, $E(t)$, $I(t)$, and $R(t)$ counts of susceptible, exposed, infected, and recovered individuals, respectively, at time $t$, we can write the likelihood function as function of the parameter vector $\theta=(a_0, a_1, b_0, b_1)$ as the individual probabilities of infection as follows:

\begin{equation}
\Lik(\theta|\Sus, \Exp, \Inf, \Rec)=P(\Sus, \Exp, \Inf, \Rec|\theta)=\prod\limits_{t}f_{t}(S(t), E(t), I(t), R(t)|\theta)
\end{equation}

where $\Sus=\{S(t)\}_t$, $\Exp=\{E(t)\}_t$, $\Inf=\{I(t)\}_t$, $\Rec=\{R(t)\}_t$, and the joint probability of all new infections occurring in time interval [t, t+1):

\begin{equation}
f_{t}(S(t), E(t), I(t), R(t)|\theta)=\Big[ \prod\limits_{i \in E(t+1)-E(t)} P(i, t) \Big] \Big[ \prod\limits_{i \in S(t+1)} (1-P(i, t)) \Big]
\end{equation}

MLE via Metropolis-Hastings MCMC requires the recalculation of $\Lik(\theta|\Sus, \Exp, \Inf, \Rec)$ a very large number of times, with varying $\theta$, in order to maximize the likelihood. In each recalculation of the likelihood, the products over all the individuals $i$ have to be evaluated, which can become prohibitive even for relatively small populations. 

An alternative MLE approach integrates compartmental models with Partially Observed Markov Process (POMP) models \cite{king2016}. Compartmental models simplify the mathematical modeling of infectious disease; however, they assume access to fully observed disease data. In reality, not all COVID-19 cases are reported, and there are several reports of infectious asymptomatic/pre-symptomatic carriers \cite{Aguirre-Duarte2020}, with some studies \cite{Nishiura2020} suggesting at least 30\% of asymptomatic cases. POMP models allow us to address such limitations by combining the simplicity of compartmental models with a probabilistic framework for the unobserved data. 

POMP models represent data $y_1^*,...,y_N^*$ collected at times $t_1<...<t_N$ as noisy, incomplete observations of an unobserved Markov process ${X(t),t \geq t_0}$. Disease transmission, represented by compartmental models, is a Markov process because the number of infectious people at time t is solely determined by the number of infectious people at time $t-\delta$. A POMP model is characterized by the transition density and measurement density of its stochastic processes. The one-step transition density is represented by $f_{X_n |X_{n-1}}(x_n | x_{n-1};\theta)$, since ${X(t)}$ is Markovian and only relies on the previous state. Meanwhile, the measurement density depends on only the state of the Markov process at that time and so is represented by $f_{Y_n |X_n}(y_n | x_n;\theta)$, where $Y_n$ is a random variable modeling the observation at time $t_n$. Hence, the entire joint density for a POMP model, including the initial density $f_{X_0}(x_0;\theta)$, is:

\begin{center}
$f_{X_{0:N}Y_{1:N}}(x_{0:N},y_{1:N};\theta)=f_{X_0}(x_0;\theta) \prod\limits_{n=1}^{N} f_{X_n |X_{n-1}}(x_n | x_{n-1};\theta)f_{Y_n |X_n}(y_n | x_n;\theta)$
\end{center}

and the marginal density for the sequence of measurements, $Y_{1:N}$, evaluated at the data, $y_{1:N}^*$, is

\begin{center}
$f_{Y_{1:N}}(y_(1:N)^*;\theta)=\int \! f_{X_{0:N}Y_{1:N}}(x_{0:N},y_{1:N};\theta) \, \mathrm{d}x_{0:N}$
\end{center}

Here the state variable $X_t$ is the vector $(S(t), E(t), I(r), R(t))$ described before. Our novel approach here will be to relate the population-level parameters in a SEIR model for COVID-19 \cite{Wang2020} with average estimations calculated over a suitable sample of individuals, which will be expressed in terms of the individual-level probabilities defined by equation \eqref{eq:10} and, ultimately, as a function of the individual-level parameters $\Theta=(a_0, a_1, b_0, b_1, \epsilon)$. In this way, our method can be seen a form of hierarchical maximum likelihood where estimation of individual-level is performed simultaneously with the population-level parameters \cite{Rouder2005}, which has the advantage of reducing variability in the recovered parameters \cite{Farrell2008}.

\subsection{SEIR model setup}

We constructed the two components of a POMP model: the unobserved process model and the measurement model. The process model, defined as a SEIR model, provides the change in true incidence of COVID at every time point, while the measurement model incorporates the fact that not all cases are observed or reported.

The underlying dynamics of COVID can be captured by a stochastic SEIR model. Most of the assumptions of a basic SEIR model  are still the same in a stochastic version. However, we add parameters that induce random fluctuations into the population and change the compartments’ rates of transfer in response to interventions. We do this by using probabilistic densities for the transition of state variables. Moreover, although disease dynamics are technically a continuous Markov process, this is computationally complex and inefficient to model, and so we make discretized approximations by updating the state variables after a time step, $\delta$. The system of discretized equations is shown below, where $B(t)$ is the number of susceptible individuals who become exposed to COVID, $C(t)$ is the number of newly infectious cases, and $D(t)$ is the number of cases that are removed from the population, all during the time step $\delta$:

\begin{equation}
\begin{aligned}
& S(t+\delta) = S(t)- B(t) \\
& E(t+\delta) = E(t)+ B(t)- C(t) \\
& I(t+\delta) = I(t)+ C(t)- D(t) \\
& R(t+\delta) = R(t)+D(t) \\
& S(t)+ E(t)+ I(t)+ R(t) = N  \\
\end{aligned}
\end{equation}

This equation describes how the sizes of the four compartments (susceptible, exposed, infectious, and removed) change between $(t,t+\delta)$. The model further assumes that the population size $N$ remains constant at every time point. We added inherent randomness to our model by setting $B(t)$, $C(t)$, and $D(t)$ as binomials. If we assume that the length of time an individual spends in a compartment is exponentially distributed with some compartment-specific rate $x(t)$, then the probability of remaining in that compartment for an additional day is $\exp(-x(t))$ and the probability of leaving that compartment is $1-\exp(-x(t))$:

\begin{equation}
\begin{aligned}
& B(t) \sim Binomial(S(t), 1-\exp(-\lambda(t))) \\
& C(t) \sim Binomial(E(t), 1-\exp(-\sigma)) \\
& D(t) \sim Binomial(I(t), 1-\exp(-\gamma)) \\
\end{aligned}
\end{equation}

The force of infection, $\lambda(t)$, is the transition rate between the susceptible and exposed classes and can be written as $\lambda(t)=\beta(t)\frac{I(t)}{N}$, where $\beta(t)$ represents the transmission rate of the disease. Furthermore, $\sigma$ is the transition rate between the exposed and infectious classes, and $\gamma$ is the transition rate between the infectious and removed compartments. $\sigma^{-1}$ represents the mean length of time a person stays in the latent stage and  $\gamma^{-1}$ represents the mean length of time a person is infectious before being removed from the population (either because of intervention efforts or natural recovery). We assume these two parameters to be constant over the course of the epidemic.

The transmission rate $\beta(t)$ can be estimated from sample averages calculated over individuals. If we recall the ILM formalism from the previous section, we can write the probability of infection of susceptible $i$ by infected contact $j$ as follows:

\begin{equation}
p_{i,j} =1-\exp\Big( -\big[a_0 + a_1 X_i] \big[b_0 + b_1 Y_j]\Big)
\end{equation}

Therefore, the transmission rate for individual $j$ is the sum of these probabilities over all the contacts:

\begin{equation}
\beta_j = \sum\limits_{i \in C(j)}  1-\exp\Big( -\big[a_0 + a_1 X_1(i)] \big[b_0 + b_1 Y_1(j)]\Big)
\end{equation}

If we are considering infected individuals from a random sample $J$  we can then estimate the transmission rate as:

\begin{equation}
\hat{\beta} = \frac{1}{|J|} \sum\limits_{j \in J}\sum\limits_{i \in C(j)}  1-\exp\Big( -\big[a_0 + a_1 X_1(i)] \big[b_0 + b_1 Y_1(j)]\Big)
\end{equation}

Given a fixed sample $J$, we can consider $\hat{\beta}=\hat{\beta}(a_0, a_1, b_0, b_1)$, that is, a function solely of the individual-level susceptibility and transmissibility coefficients.

Although it is impossible to directly record the number of people that are susceptible, exposed, infectious, and removed directly, the publicly available data tells us the number of observed cases per day. The mean number of observed cases per day is the true number of cases multiplied by the reporting rate $\rho$ ($\rho<1$). This can be modeled with a binonial distribution of parameters $C(t)$ and $\rho$:

\begin{equation}
y_t|C(t) \sim Binomial(C(t), \rho)
\end{equation}

The process and measurement models define our final POMP model. For each time point, the process model generates the number of new cases based on binomial distributed counts. The measurement model then estimates the observed number of cases based on the true number of cases and reporting rate. 

\subsection{Iterated filtering method}

The likelihood function for the POMP models is the density function evaluated with data at a candidate set of parameter values. It is computationally simpler to work with the log likelihood, $l(\theta)=\log f(y_{1:N};\theta)$, so that we can deal with sums instead of products. We used a simulation-based approach to avoid solving the density function analytically, in which we simulated the random variable $Y_{1:N}$, which implicitly defines the density function. Likelihood evaluation via Sequential Monte Carlo (SMC) techniques is one standard method to obtain the log likelihood for POMP models, because it simulates sample paths rather than requiring explicit transition probabilities. Exploiting the Markov property of the process, it is possible to use these paths to sample the parameter space much more efficiently than with regular MCMC, thanks to the iterated filtering method.

We factorized the likelihood as the product of conditional likelihoods:

\begin{equation}
L(\theta)=\prod\limits_{n=1}^N L_{n|1:n-1}(\theta)
\end{equation}

where $L_{n|1:n-1}(\theta)=P[y_n^*|y_{1:n-1}^*;\theta]$ and there are $N$ time points. The structure of a POMP model then implies the representation of $L_{n|1:n-1}(\theta)$ as

\begin{equation}
L_{n|1:n-1}(\theta)=\int \! P[y_n^*|x_n;\theta]P[x_n|y_{1:n-1}^*;\theta] \, \mathrm{d}x_n
\end{equation}

so that the final expression for the likelihood is: 

\begin{equation}
\prod\limits_{n=1}^N \int \! P[y_n^*|x_n;\theta]P[x_n|y_{1:n-1}^*;\theta] \, \mathrm{d}x_n
\end{equation}

In this last equation, although $P[y_n^*|x_n;\theta]$ is simple to calculate (using the observation process), $P[x_n|y_{1:n-1}^*;\theta]$ is more difficult to evaluate. We can use the Markov property to determine an expression for this probability, known as the prediction formula: 

\begin{equation}
P[x_n|y_{1:n-1}^*;\theta]=\int \! P[x_n|x_{n-1};\theta]P[x_{n-1}|y_{1:n-1}^*;\theta] \, \mathrm{d}x_{n-1}
\end{equation}

We can then use Bayes’ Theorem to determine an expression for $P[x_{n-1}|y_{1:n-1}^*;\theta]$, known as the filtering formula: 

\begin{equation}
P[x_n|y_{1:n}^*;\theta]=P[x_n|y_n^*,y_{1:n-1}^*;\theta]=\frac{P[y_n^*|x_n;\theta]P[x_n|y_{1:n-1}^*;\theta]}{\int \! P[y_n^*|x_n;\theta]P[x_n|y_{1:n-1}^*;\theta] \, \mathrm{d}x_n}
\end{equation}

The prediction and filtering formulas give us a recursion. Specifically, the prediction formula calculates the prediction distribution at time $t_n$, $f_(X_n|Y_{1:n-1})(x_n|y_{1:n-1}^*)$, at time $t_n$ by using the filtering distribution at time $t_{n-1}$, $f_(X_{n-1}|Y_{1:n-1})(x_{n-1}|y_{1:n-1}^*)$,at time $t_{n-1}$ . Meanwhile, the filtering formula gives us the filtering distribution at time $t_n$ using the prediction distribution at time $t_n$. 

In SMC, we use Monte Carlo techniques to sequentially estimate the integrals in the prediction and filtering recursions, which in turn allows us to estimate $P[x_n|y_{1:n-1}^*;\theta]$. This is done by generating a swarm of $J$ particles that are propagated forward based on the process model and then filtered and altered to fit the next data point more closely. Because of this, SMC is commonly known as the particle filter.

Iterated filtering \cite{Ionides2006} allows to more efficiently obtain MLEs of parameters in partially observed dynamical systems, such as POMPs. It works by defining a set of initial values for the parameter vector $\theta$ and a fixed number of iterations, $M$. For every iteration, we apply a basic particle filter (Equation 9 above) to the model and add stochastic perturbations to the parameters so that they take a random walk through time. At the end of the time series, we use the final value of the parameters as the starting point for the next iteration but reducing (“cooling”) the random walk variance. After completing the $M$ iterations, we obtain the Monte Carlo maximum likelihood estimate, $\theta_M$, and its corresponding log likelihood. In contrast, Monte Carlo likelihood by direct simulation scales poorly with dimension. It requires a Monte Carlo effort that scales exponentially with the length of the time series, and so is infeasible on anything but a short data set.

\section{Simulation study}

\subsection{Parameter estimation}

Millions of individuals worldwide have self-reported symptoms associated with COVID-19 infection through numerous websites and apps specifically designed for that purpose \cite{Menni2020}. Meanwhile, anonymized mobility data generated by cellphones has been aggregated from several sources and made available for research \cite{covid19mobility}. However, this currently available data is not enough to evaluate the ILMs described above. These models can potentially provide infection risk predictions aggregating several sources of health and epidemiological data from individuals, including symptoms, demographics, and contact information. Approaches incorporating this kind of data, collected through contact tracing and symptom reporting apps, have been proposed recently by several groups \cite{sphinxteam, Alsdurf2020}, and have lead to consider the privacy risks presented by this data and possible mitigation of those risks. Here, we are focusing primarily on the parameter estimation problem, assuming that it is possible to acquire the data securely, but will make some observations regarding privacy in the conclusions.

Since the detailed data needed to calculate our ILMs is not currently available, we have started by conducting a purely computational study where the individual-level data is generated by means of  agent-based models (ABMs). These models allow us to simulate behaviours of individuals in a large population, and obtain data that mirrors what we could collect with contact tracing and symptom reporting apps in real life. One advantage of using ABMs at this stage is that we can define the ground truth of the ILM by specifying the coefficients in the susceptibility and transmissibiliy functions, this allow us to evaluate the accuracy of our parameter estimation methods. 

For the purpose of running the ABMs simulations, we used the COMOKIT COVID-19 SEIR model \cite{Drogoul2020} implemented in the GAMA software \cite{Taillandier2019}, a general ABM simulator allowing for a wide range of options through a custom modeling language and supporting GIS layers to represent specific geographical in detail. In particular, we run simulations using a scenario of a COVID-19 outbreak in Vietnam without any containment strategies, with a population of nearly 10,000 individuals. We adapted the SEIR model provided in COMOKIT to incorporate the individual probabilities of infection as defined by Formula \eqref{eq:10}, with a number of different selections of parameters $\theta=(a_0, a_1, b_0, b_1)$. We arbitrarily defined three sets of ground truth parameters, $\theta_1=(a_0=0.2, a_1=2.0, b_0=0.2, b_1=2.0)$, $\theta_2=(a_0=0.5, a_1=1.5, b_0=0.5, b_1=0.8)$, and $\theta_3=(a_0=1.0, a_1=2.0, b_0=1.0, b_1=1.0)$. Given the symmetric form of \eqref{eq:10}, we expanded the product of the transmissibility and susceptibility functions to arrive to the following reparametrization: 

\begin{equation}
\hat{\beta} = \frac{1}{|J|} \sum\limits_{j \in J}\sum\limits_{i \in C(j)}  1-\exp\Big(- \big[c_{00} + c_{01} Y_1(j) + c_{10} X_1(i) + c_{11} X_1(i) Y_1(j)]\Big)
\end{equation}

The relationship between the original and new parameters is given by:

\begin{equation}
\begin{aligned}
& a_0 b_0= c_{00} \\
& a_0 b_1= c_{01} \\
& a_1 b_0= c_{10} \\
& a_1 b_1= c_{11}
\end{aligned}
\end{equation}

from which we can express the ratios $a_0/a_1$ and $b_0/b_1$ as a function of the new parameters $c_{00}$, $c_{01}$, $c_{10}$, and $c_{11}$:

\begin{equation}
\begin{aligned}
& \frac{a_0}{a_1} = \frac{c_{01}}{c_{11}} = \frac{c_{00}}{c_{10}} \\
& \frac{b_0}{b_1} = \frac{c_{00}}{c_{01}} = \frac{c_{10}}{c_{11}} \\
\end{aligned}
\label{eq:eq2}
\end{equation}

\begin{figure} 
  \centering
  \includegraphics[scale=0.4]{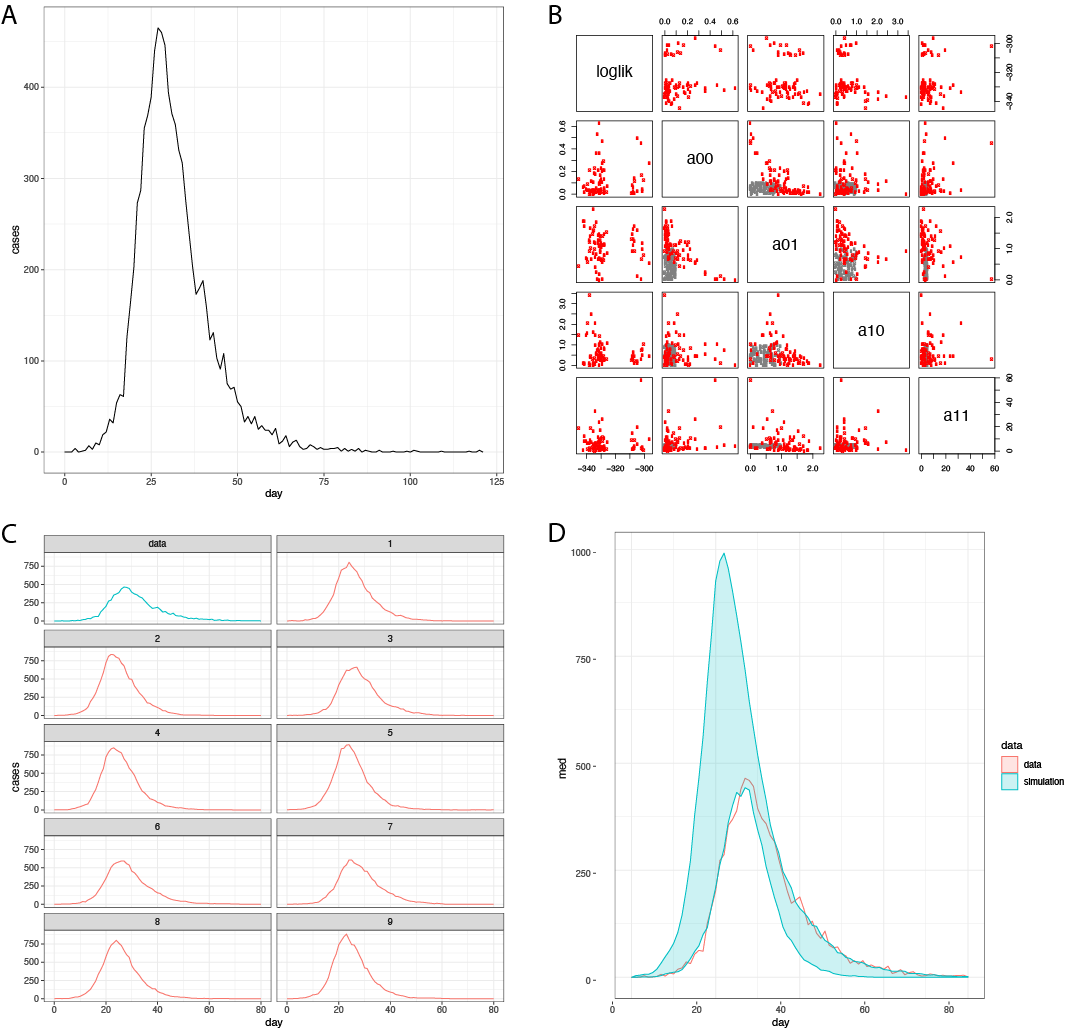}
  \caption{Result of MLE with POMP on synthetic data generated with GAMA: (A) Curve of new infected cases over time. (B) Scatterplot matrix showing all the initial points (gray) and final (red) points in parameter space from the MLE runs. (C) GAMA curve (blue) and 9 simulated curves (red) generated with the MLE parameters. (D) Range covered between the 5 and 95 percentiles from 100 simulated curves using the MLE parameters. }
  \label{fig:fig1}
\end{figure}

We assume the mean latent and infectious times to be known, and we can estimate them from the GAMA data. In the case of the COMOKIT simulations, have $\sigma=0.26$ and $\gamma=0.6$, both in units of 1/day. Therefore, the only parameters in the SEIR model at the population level are $c_{00}$, $c_{01}$, $c_{10}$, and $c_{11}$, which we estimate by applying iterated filtering with POMP. An MLE run corresponding to the the underlying parameters $\theta_1$ is shown in figure \ref{fig:fig1}.

\begin{figure} 
  \centering
  \includegraphics[scale=0.4]{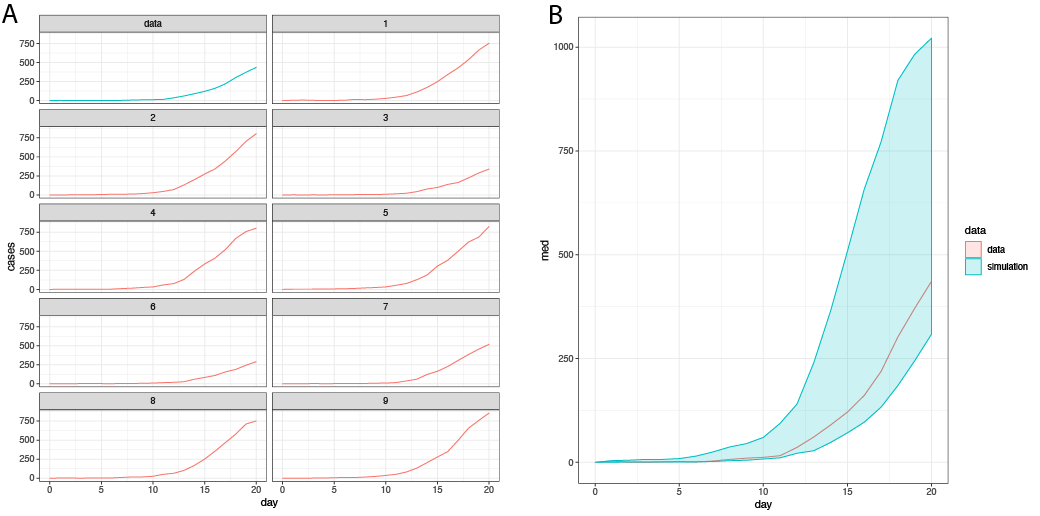}
  \caption{Range of synthetic epidemic data used to fit the parameters. (A) GAMA curve (blue) and 9 simulated curves (red) generated with the resulting MLE parameters. (B) Range covered between the 5 and 95 percentiles from 100 simulated curves using the MLE parameters. }
  \label{fig:fig2}
\end{figure}

An issue we encountered with the first round of MLE runs is that, as the ABM simulation progressed, the compartment of susceptible individuals gets depleted as more become infected, and so the $\hat{\beta}$ estimator becomes increasingly biased. In order to account for this problem, we fit the GAMA data only for the initial stages of the simulated epidemic, when the number of new infectious cases is still increasing due to the large percentage of susceptible individuals. The range of the data with enough susceptibles is shown in figure \ref{fig:fig2}.

\begin{figure} 
  \centering
  \includegraphics[scale=0.5]{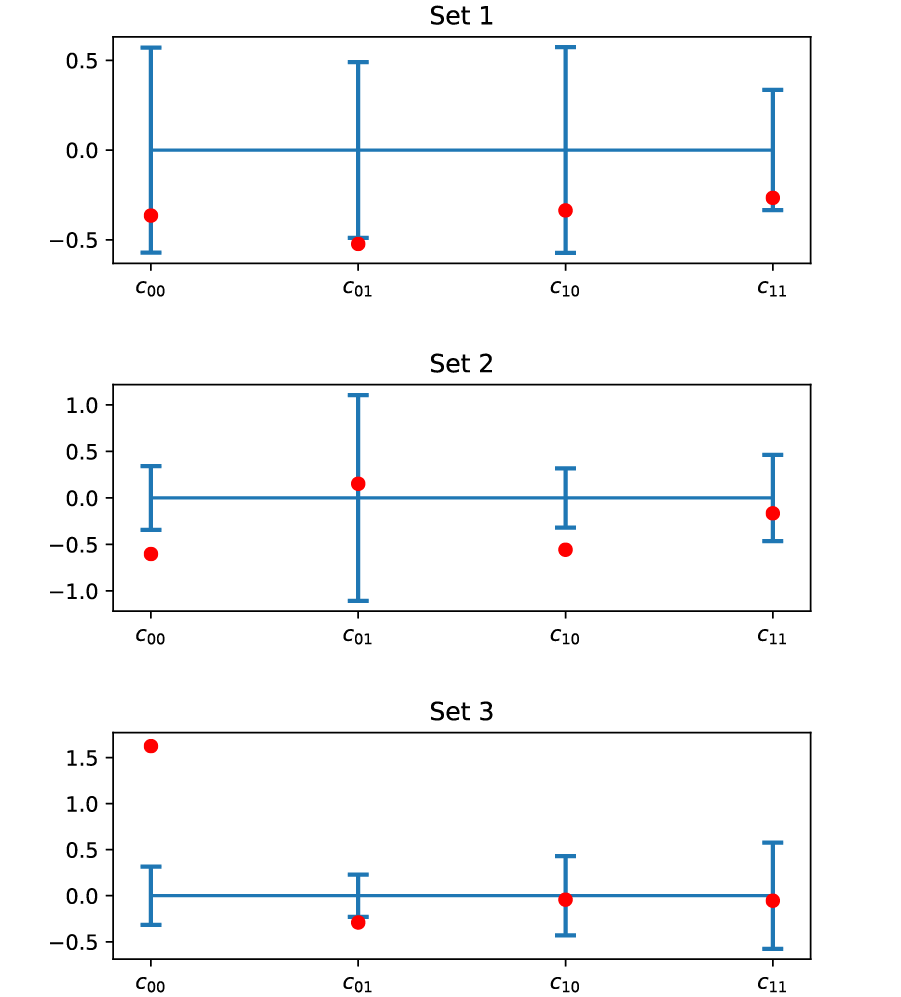}
  \caption{Plots representing the standard deviation of the top 10 MLEs for each parameter as an error bar. The red circle indicates the true value of the parameter. All values are scaled to the mean MLE for each parameter.}
  \label{fig:fig3}
\end{figure}

The ground truth values for the $c_{00}$, $c_{01}$, $c_{10}$, and $c_{11}$ parameters, and the mean and standard deviation over the 10 highest likelihoods are listed in table \ref{tab:table1}. Most of the true values fall within a standard deviation from the mean MLE, as seen in \ref{fig:fig3}, with the exception of $c_{00}$ in the parameter set 3.

\begin{table}
\caption{Maximum likelihood estimates for all the parameter sets}
\centering
\begin{tabular}{llll}
    \rowcolor[rgb]{0.796,0.796,0.796} \multicolumn{1}{c}{} & \multicolumn{3}{c}{Set 1}         \\
    \rowcolor[rgb]{0.878,0.878,0.878}                      & True value & MLE mean & MLE sdev  \\
    {\cellcolor[rgb]{0.878,0.878,0.878}}$c_{00}$            & 0.04       & 0.063    & 0.036     \\
    {\cellcolor[rgb]{0.878,0.878,0.878}}$c_{01}$            & 0.40       & 0.839    & 0.411     \\
    {\cellcolor[rgb]{0.878,0.878,0.878}}$c_{10}$            & 0.40       & 0.603    & 0.346     \\
    {\cellcolor[rgb]{0.878,0.878,0.878}}$c_{11}$            & 4.00       & 5.452    & 1.825     \\
    \rowcolor[rgb]{0.796,0.796,0.796}                      & \multicolumn{3}{c}{Set 2}         \\
    \rowcolor[rgb]{0.878,0.878,0.878}                      & True value & MLE mean & MLE sdev  \\
    {\cellcolor[rgb]{0.878,0.878,0.878}}$c_{00}$            & 0.25       & 0.629    & 0.216     \\
    {\cellcolor[rgb]{0.878,0.878,0.878}}$c_{01}$            & 0.40       & 0.347    & 0.384     \\
    {\cellcolor[rgb]{0.878,0.878,0.878}}$c_{10}$            & 0.75       & 1.690    & 0.538     \\
    {\cellcolor[rgb]{0.878,0.878,0.878}}$c_{11}$            & 1.20       & 1.438    & 0.667     \\
    \rowcolor[rgb]{0.796,0.796,0.796}                      & \multicolumn{3}{c}{Set 3}         \\
    \rowcolor[rgb]{0.878,0.878,0.878}                      & True value & MLE mean & MLE sdev  \\
    {\cellcolor[rgb]{0.878,0.878,0.878}}$c_{00}$            & 1.0        & 0.381    & 0.120     \\
    {\cellcolor[rgb]{0.878,0.878,0.878}}$c_{01}$            & 1.0        & 1.409    & 0.322     \\
    {\cellcolor[rgb]{0.878,0.878,0.878}}$c_{10}$            & 2.0        & 2.088    & 0.901     \\
    {\cellcolor[rgb]{0.878,0.878,0.878}}$c_{11}$            & 2.0        & 2.113    & 1.218    
\end{tabular}
\label{tab:table1}
\end{table}

In order to recover the original parameters $a_0$, $a_1$, $b_0$, and $b_1$ that are needed in the susceptibility and transmisibility functions, we can use the ratios in equation \ref{eq:eq2}. An initial calculation simply taking the top estimates of the $c_{ij}$'s to calculate the mean and standard deviation of the $a$'s and $b$'s gave values with high errors when compared with the true rations, which seems to be caused by fluctuations in the $c_{ij}$'s. A better approximation to the ratios is given by this heuristic formula:

\begin{equation}
\begin{aligned}
& \frac{a_0}{a_1}=0.5\frac{\overline{MLE}(c_{01})}{\overline{MLE}(c_{11})} + 0.5\frac{\overline{MLE}(c_{00})}{\overline{MLE}(c_{10})} \\
& \frac{b_0}{b_1}=0.5\frac{\overline{MLE}(c_{10})}{\overline{MLE}(c_{11})} + 0.5\frac{\overline{MLE}(c_{00})}{\overline{MLE}(c_{01})}
\end{aligned}
\label{eq:eq3}
\end{equation}

where $\overline{MLE}(c_{ij})$ represents the mean of the parameter $c_{ij}$ taken over the 10 top MLEs. This formula smooths out the fluctuations in the individual parameters, and gives a better result, shown in table \ref{tab:table2}.

\begin{table}
\caption{Approximation to the original parameter ratios using MLEs}
\centering
\begin{tabular}{lll}
\rowcolor[rgb]{0.796,0.796,0.796}         & \multicolumn{2}{c}{Set 1}  \\
\rowcolor[rgb]{0.878,0.878,0.878}         & True ratio & MLE approx.   \\
{\cellcolor[rgb]{0.878,0.878,0.878}}$a_0/a_1$ & 0.10       & 0.13          \\
{\cellcolor[rgb]{0.878,0.878,0.878}}$b_0/b_1$ & 0.10       & 0.09          \\
\rowcolor[rgb]{0.796,0.796,0.796}         & \multicolumn{2}{c}{Set 2}  \\
\rowcolor[rgb]{0.878,0.878,0.878}         & True ratio & MLE approx.   \\
{\cellcolor[rgb]{0.878,0.878,0.878}}$a_0/a_1$ & 0.33       & 0.31          \\
{\cellcolor[rgb]{0.878,0.878,0.878}}$b_0/b_1$ & 0.62       & 1.50          \\
\rowcolor[rgb]{0.796,0.796,0.796}         & \multicolumn{2}{c}{Set 3}  \\
\rowcolor[rgb]{0.878,0.878,0.878}         & True ratio & MLE approx.   \\
{\cellcolor[rgb]{0.878,0.878,0.878}}$a_0/a_1$ & 0.5        & 0.42          \\
{\cellcolor[rgb]{0.878,0.878,0.878}}$b_0/b_1$ & 1.0        & 0.63         
\end{tabular}
\label{tab:table2}
\end{table}

\subsection{Risk calculation}

Once the parameters of the model have been determined through MLE using POMP, in particular the individual-level coefficients ($c_{00}$, $c_{01}$, $c_{10}$, $c_{11}$), we can compute individual risks of infection using equation \eqref{eq:14}. We use the ABM in GAMA with the three parameter sets and a random assignment of symptoms for susceptible and infected individuals using the symptom prevalence for the US listed in \cite{Menni2020}. Instead of using the ground truth individual-level parameters, we generate random perturbation of $a_0$ and $b_0$, and then obtained $a_1$ and $b_1$ with the ratio estimates in \eqref{eq:eq3}.

As a fist sanity check, we calculated the mean risk for susceptible and infected individuals over entire simulation parameter runs with each parameter set, and we obtained the results shown in table \ref{tab:table3}. Difference between the risks for susceptible and infected is very small for parameter set 3, and the reason for this is that this set yields higher probabilities of infection across all individuals, resulting in more uniform risk score values. We then make predictions of infected status for all agents in each simulation based on their risk, using the mean infected risk minus the standard deviation, as listed in \ref{tab:table3}, to calculate the threshold to predict infection.

\begin{table}
\caption{Mean and standard deviation of individual risks for susceptible ($R_S$) and infected individuals ($R_I$)}
\centering
\begin{tabular}{lll}
\rowcolor[rgb]{0.878,0.878,0.878}         & $R_S$        & $R_I$         \\
{\cellcolor[rgb]{0.878,0.878,0.878}}Set 1 & 0.24$\pm$0.25 & 0.52$\pm$0.34  \\
{\cellcolor[rgb]{0.878,0.878,0.878}}Set 2 & 0.28$\pm$0.24 & 0.45$\pm$0.32  \\
{\cellcolor[rgb]{0.878,0.878,0.878}}Set 3 & 0.55$\pm$0.31 & 0.63$\pm$0.31 
\end{tabular}
\label{tab:table3}
\end{table}

\begin{figure} 
  \centering
  \includegraphics[scale=0.8]{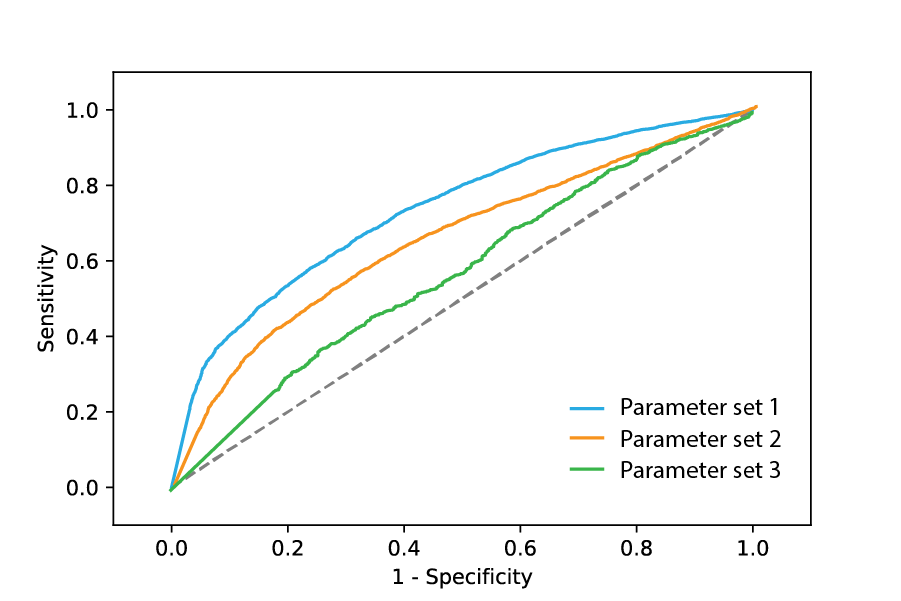}
  \caption{Receiver Operating Characteristic (ROC) Curve for the infection predictor based on equation \eqref{eq:14}, generated for the 3 parameter sets considered in the paper}
  \label{fig:fig4}
\end{figure}

The area under the receiver characteristic curve (AUC), the 95\% confidence interval (CI) for the AUC values, sensitivity, specificity, and overall accuracy of the infection prediction for each parameter set are listed in \ref{tab:table4}. The corresponding receiver characteristic curves are plotted in \ref{fig:fig4}. We can see how the predictions fare for each parameter set, and set 1 has the highest AUC and best balance of sensitivity and specificity. Even though set 2 and 3 exhibit higher sensitivity, their specificity is fairly low, meaning that in those parameter sets, the risk predictor results in many false positives. In terms of using the risk to decide when to quarantine agents, a high false positive rate (agents that are not infected get needlessly quarantined) is arguably better than a high false negative rate (infected agents are not quarantines and go on to transmit the pathogen). 

\begin{table}
\caption{Performance measures for the infection predictor based on equation \eqref{eq:14}, calculated on the 3 parameters sets considered in the paper}
\centering
\begin{tabular}{lllll}
\rowcolor[rgb]{0.878,0.878,0.878}         & AUC (95\% CI)     & Accuracy & Sensitivity & Specificity  \\
{\cellcolor[rgb]{0.878,0.878,0.878}}Set 1 & 0.74 (0.73, 0.74) & 0.65     & 0.78        & 0.53         \\
{\cellcolor[rgb]{0.878,0.878,0.878}}Set 2 & 0.65 (0.64, 0.66) & 0.62     & 0.83        & 0.29         \\
{\cellcolor[rgb]{0.878,0.878,0.878}}Set 3 & 0.57 (0.56, 0.59) & 0.76     & 0.81        & 0.28   
\end{tabular}
\label{tab:table4}
\end{table}

\begin{figure} 
  \centering
  \includegraphics[scale=0.4]{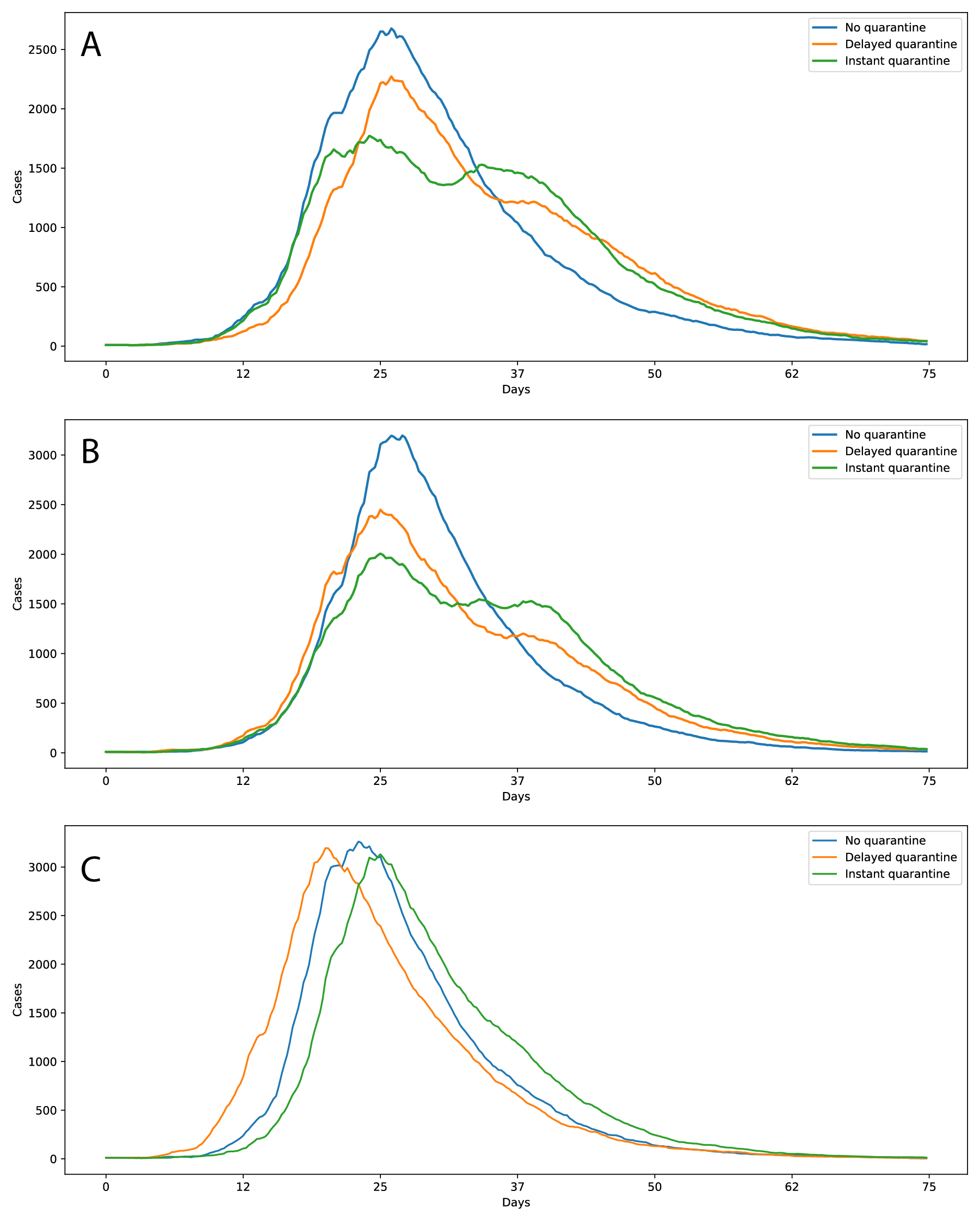}
  \caption{Plot of new cases for the three parameter sets (A: set 1, B: set 2, C: set 3), under three intervention scenarios: no quarantine (blue curve), risk-based quarantine with 4 day delay (orange curve), and risk-based quarantine with instant availability of infectious status of contacts.}
  \label{fig:fig5}
\end{figure}

Finally, we run ABM simulations where the risk values where used to quarantine for 14 days those individuals with a risk higher than a given threshold. We simulated two scenarios, in the first scenario, there was a delay of 4 days between the risk calculation and its use to determine quarantine, in order to model the fact that the infectious status of contacts is not determined instantaneously, but with a lag caused by the symptom onset time (and also by the delay in obtaining test results). In the second scenario, the risk was updated instantly with the information of the infected contacts, this represents an unrealistic situation where infectious status is known upon interaction but gives an upper bound for the performance of the intervention. The results form these simulations are shown in figure \ref{fig:fig5}. The epidemic curves of new cases for the three scenarios (no quarantine, delayed quarantine, and instant quarantine) suggest that risk-based quarantine could help in lowering the peak of new cases and spread them over a longer period of time (e.g.: "flattening the curve"), with the most pronounced effect in the case of instantaneous availability of infection status, which is impossible in practice but providing a "maximum curve flattening".

\section{Conclusions}

We constructed an statistical inference framework that allows to obtain individual-level epidemic parameters by applying MLE to population-level case data. We tested this framework using an agent-based model to generate epidemic data resolved at the individual level. As part of this framework, we defined an individual-level epidemic risk model that depends on data such as demographics, medical condition, self-reported symptoms, and contact tracing information. These models could be trained on aggregate data provided by consenting users of a mobile app, for example, and then evaluated locally by the rest of the users. These models could also incorporate additional data, such as spacial random effects. The initial simulation experiments are promising and suggest that is possible to: (1) obtain good estimates for the individual-level parameters by applying MLE on the population level data, (2)  predict who is infected and who is not using the individual-level risks, and (3) carry out interventions based on the individual-level risks, such as quarantine, that could help in lowering the peak of the epidemic, i.e.: “flattening the curve”. However, this work in in its initial stages and has several limitations. First of all, the individual-level models we considered so far are very simple, including only two somewhat artificial covariates (immune and symptomatic levels). Secondly, data to train the models was obtained completely from simulated experiments, thus, we need to extend to and validate on real data. Third, our risk calculation requires knowledge on confirmed cases in order to determine exposure events, which might not be readily available or accessible. Other approaches \cite{sphinxteam, Alsdurf2020} are based on estimating the probabilities of all possible states an individual can be in (susceptible, infected, recovered) based on the available information (symptoms, tests, contacts, etc) and then having this information be shared across the individuals through a mobile app in order to update the probabilities as new information is obtained. Our approach has the advantage of being simpler, but could also incorporate some of these ideas to lift the requirement of exact infectious states to be known in advance to the calculation of the risks score. We envision the computational framework presented in this work as the basis for a system that could be used to estimate risks of infection for diseases other than COVID-19.

\section{Acknowledgments}
The authors would like to thank Aaron King for discussions about the model and revision of the POMP code, Hayden Metsky for suggestions regarding MLE, Brandon Westover for comments about method evaluation, and members of the Machine Learning and Optimization Lab at EPFL for support and feedback.

\section{Availability of code and data}
All the GAMA parameters and R scripts are available under the MIT license at this repository: https://github.com/broadinstitute/ILM-COVID19-risk, together with the simulated data used in the analysis.

\printbibliography

\end{document}